# Photoinduced Magnetization in a Thin Fe-CN-Co Film

J.-H. Park,[a,*] Y.D. Huh,[b,c] E. Čižmár,[a] S.J. Gamble,[a] D.R. Talham,[b] and M.W. Meisel[a]

[a]Department of Physics and Center for Condensed Matter Sciences, University of Florida, Gainesville, FL 32611-8440, U.S.A.

[b]Department of Chemistry, University of Florida, Gainesville, FL 32611-7200, U.S.A.

[c]Department of Chemistry, Dankook University, Seoul 140-714, Korea.

**Abstract**

The magnetization of a thin Fe-Co cyanide film has been investigated from 5 K to 300 K and in fields up to 500 G. Upon illumination with visible light, the magnetization of the film rapidly increases. The original cluster glass behavior is further developed in the photoinduced state and shows substantial changes in critical temperature and freezing temperature. © 2001 Elsevier Science. All rights reserved

Since Sato and co-workers reported the optical switching of the magnetic properties of Fe-Co cyanide [1], numerous theoretical and experimental studies have focused on this type of photoinduced magnetic phenomena [2, 3]. Here we report the photoinduced magnetization of Fe-Co cyanide with Rb ions in the form of a unique thin film.

Thin film samples were prepared by first coating a hydrophobic surface with a two-dimensional Fe-CN-Co network prepared as a monolayer at the air-water interface. This well structured 2D monolayer was transferred using the Langmuir-Blodgett (LB) technique and served as a template for future bulk film growth [4]. Subsequent immersions of the resulting film in an aqueous solution of hexacyanoferrate, and into a separate solution of $Rb^+$ and $Co^{2+}$ ions, completed one deposition cycle of the Fe-CN-Co network. For the magnetic studies, 50 or 150 deposition cycles were used. The resulting transparent film was cut into small rectangles and stacked into a gel-cap with the surface parallel to the direction of the applied magnetic field. Two optical fibers (diameter = 250 μm) were then introduced to the stacked films, and then connected to room-temperature light sources using a homemade probe. The magnetization of the samples was investigated using a Quantum Design MPMS XL SQUID magnetometer, while the visible light was provided by a Halogen lamp or by LEDs of various colors.

In an external magnetic field of 50 G, a sample was cooled to 5 K and was monitored as a function of time. After white light began to continuously illuminate the specimen, the magnetization increased more than 300% within 30 minutes and slowly tended towards saturation, see Fig. 1. The magnetization of the film was activated by excitation at different visible frequencies, see inset of Fig.1. To date, we have not observed any decrease of the magnetization from its photoinduced state while illuminating with infrared. The mechanism of this photoinduced magnetism can be explained by considering an optically induced charge transfer between the diamagnetic $Fe^{2+}$-CN-$Co^{3+}$ state and the paramagnetic $Fe^{3+}$-CN-$Co^{2+}$ state [1, 2].

Figure 2 shows the field-cooled (fc) and the zero-field-cooled (zfc) magnetization of the film as a function of temperature with an applied field of 500 G. For the photoinduced state, the magnetization increases below a critical temperature, $T_c \approx 18$ K [5], which is indicative of spontaneous magnetization. The apparent divergence between $M_{fc}$ and $M_{zfc}$ below a temperature, $T_{irr} \approx 8$ K,

---

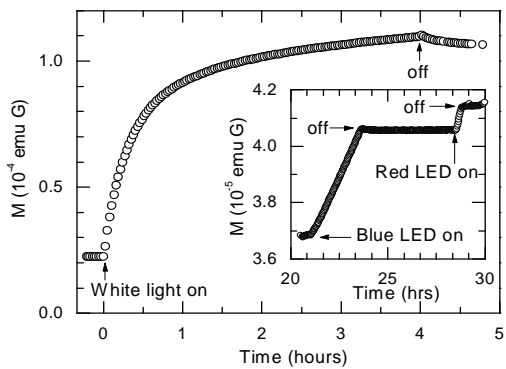

Fig. 1. Magnetization vs time at 5 K and in 50 G.

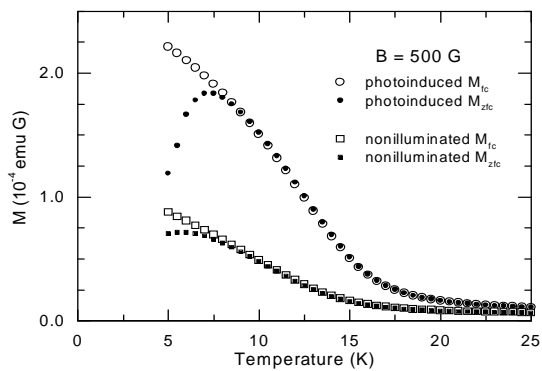

Fig. 2. Field-cooled ($M_{fc}$) and zero-field-cooled ($M_{zfc}$) magnetizations of both photoinduced and nonilluminated states of film as a function of temperature.

suggests an irreversible $M_{zfc}$ below $T_{irr}$. As the temperature is further lowered below $T_{irr}$, $M_{zfc}$ reaches a maximum, $M_{max}$, at $T_{max} \approx 7$ K. All of these features are consistent cluster spin-glass behavior [6, 7].

In the nonilluminated state, similar magnetic behavior was observed, however there are quantitative differences in the characteristic temperatures ($T_c$, $T_{irr}$, and $T_{max}$). Table 1 summarizes the characteristic temperatures of both the nonilluminated and the photoinduced states of the film. It is noteworthy that $T_c$ and $T_{max}$ are higher in the photoiduced state. These changes can be interpreted as consequences of the increased size and concentration of the magnetic clusters. Larger clusters possess a higher $T_c$ since the magnetic coherence length is less restricted by the structural dimensions. In addition, an increase in the density of the larger clusters leads to enhanced interactions between these domains, thereby raising the freezing temperature, $T_{max}$. This type of interpretation was employed by Pejaković *et al.* in their study of a similar but powder material [8], and their results are qualitatively consistent with our data and interpretations.

In summary, we have studied the photoinduced magnetization of a new low dimensional system, a thin film of Fe-Co cyanide with Rb ions. In this novel thin film preparation, high spin states of the Fe and Co interact and experience long range ordering below 18 K. The spins form domains that exhibit behavior consistent with a cluster spin-glass description. Upon illumination with light, the population of the high spin states increases, resulting in a rapid increase of the magnetization of the film and a modification of the cluster spin-glass properties.

Table 1. Characteristic temperatures in an applied field of 500 G.

|          | $T_c$ (K) | $T_{irr}$ (K) | $T_{max}$ (K) |
| -------- | --------- | ------------- | ------------- |
| Light    | 18        | 8             | 7             |
| No light | 15        | 8             | 6             |

### Acknowledgements

This work was supported, in part, by NSF DMR-9900855 (DRT), NSF DMR-0113714 (MWM and DRT), ACS-PRF 36163-AC5 (MWM and DRT), and NSF DGE-0209410 (EC). We gratefully acknowledge early contributions from J.T. Culp.